\documentclass[prd,superscriptaddress,amsfonts,amssymb,amsmath,showpacs,twocolumn]{revtex4}
\usepackage{bm}
\usepackage{amsfonts}
\usepackage{latexsym}
\usepackage[latin1]{inputenc}
\usepackage{graphicx}
\usepackage{amsmath}
\usepackage{palatino}
\usepackage{mathpazo}
\usepackage{textcomp}
\usepackage{float}
\usepackage{booktabs}
\usepackage{dcolumn}
\usepackage{hyperref}
\hypersetup{colorlinks,citecolor=blue}
\usepackage{amsmath}
\usepackage{xcolor}
\usepackage{orcidlink}
\usepackage[caption=false]{subfig}
\usepackage{commath}
\def\jnl@style{\it}
\def\aaref@jnl#1{{\jnl@style#1}}

\def\aaref@jnl#1{{\jnl@style#1}}

\def\aj{\aaref@jnl{AJ}}                   
\def\apj{\aaref@jnl{ApJ}}                 
\def\apjl{\aaref@jnl{ApJ}}                
\def\apjs{\aaref@jnl{ApJS}}               
\def\apss{\aaref@jnl{Ap\&SS}}             
\def\aap{\aaref@jnl{A\&A}}                
\def\aapr{\aaref@jnl{A\&A~Rev.}}          
\def\aaps{\aaref@jnl{A\&AS}}              
\def\mnras{\aaref@jnl{Mon.~Not.~Roy.~Astron.~Soc.}}             
\def\prd{\aaref@jnl{Phys.~Rev.~D}}        
\def\prc{\aaref@jnl{Phys.~Rev.~C}}  
\def\prl{\aaref@jnl{Phys.~Rev.~Lett.}}    
\def\qjras{\aaref@jnl{QJRAS}}             
\def\skytel{\aaref@jnl{S\&T}}             
\def\ssr{\aaref@jnl{Space~Sci.~Rev.}}     
\def\zap{\aaref@jnl{ZAp}}                 
\def\nat{\aaref@jnl{Nature}}              
\def\aplett{\aaref@jnl{Astrophys.~Lett.}} 
\def\apspr{\aaref@jnl{Astrophys.~Space~Phys.~Res.}} 
\def\physrep{\aaref@jnl{Phys.~Rep.}}      
\def\physscr{\aaref@jnl{Phys.~Scr}}       
\def\commat{\aaref@jnl{Comm.~Math.~Phys.}}              
\def\science{\aaref@jnl{Science}}               
\def\cqg{\aaref@jnl{Classical Quant.~Grav.}}            
\def\jpcs{\aaref@jnl{JPCS}}                                     
\def\ijmpd{\aaref@jnl{Int.~J.~Mod.~Phys.~D}}                    
\def\grg{\aaref@jnl{Gen.~Relat.~Gravit.}}               
\def\rpp{\aaref@jnl{Rep.~Prog.~Phys.}}          
\def\npa{\aaref@jnl{Nucl.~Phys.~A}}        
\def\lrr{\aaref@jnl{Living Rev.~Rel.}}                   
\def\jcap{\aaref@jnl{J.~Cosmology Astropart.~Phys.}}    
\def\rmp{\aaref@jnl{Rev.~Mod.~Phys.}}   
\def\epjc{\aaref@jnl{Eur.~Phys.~J.~C}}

\allowdisplaybreaks[1]

\begin{document}
\color{red}

\title{Energy Conditions in $f(Q)$ gravity}

\author{Sanjay Mandal\orcidlink{0000-0003-2570-2335}}
\email{sanjaymandal960@gmail.com}
\affiliation{Department of Mathematics, Birla Institute of Technology and
Science-Pilani,\\ Hyderabad Campus, Hyderabad-500078, India.}
\author{P.K. Sahoo\orcidlink{0000-0003-2130-8832}}
\email{pksahoo@hyderabad.bits-pilani.ac.in}
\affiliation{Department of Mathematics, Birla Institute of Technology and
Science-Pilani,\\ Hyderabad Campus, Hyderabad-500078, India.}
\author{J.R.L. Santos\orcidlink{0000-0002-9688-938X}}
\email{joaorafael@df.ufcg.edu.br}
\affiliation{UFCG - Universidade Federal de Campina Grande - Unidade Acad\^{e}mica de F\'isica,  58429-900 Campina Grande, PB, Brazil.}
\date{\today}
\begin{abstract}

A complete theory of gravity impels us to go beyond Einstein's General Relativity. One promising approach lies in a new class of teleparallel theory of gravity named $f(Q)$, where the nonmetricity $Q$ is responsible for the gravitational interaction. The important roles any of these alternative theories should obey are the energy condition constraints. Such constraints establish the compatibility of a given theory with the causal and geodesic structure of space-time. In this work, we present a complete test of energy conditions for $f(Q)$ gravity models. The energy conditions allowed us to fix our free parameters, restricting the families of $f(Q)$ models compatible with the accelerated expansion our Universe passes through. Our results straight the viability of $f(Q)$ theory, leading us close to the dawn of a complete theory for gravitation.      
\end{abstract}



\maketitle

\section{Introduction}\label{I}

The dark sector of the Universe is one of the most challenging problems that science has been facing. The fact the approximately $96$\% of the matter and energy contents of the Universe is unknown, suggests that our standard description of gravity is incomplete \cite{Capozziello_book}. The standard description is based on Einstein-Hilbert field equations for General Relativity (GR), and despite this issue on the dark sector, such a theory has been successfully tested since 1919, when it predicted the perihelion advance of Mercury. Recently, the observations of gravitational waves performed by LIGO and Virgo \cite{lv_papers}, besides the captured images of a black hole from Event Horizon Telescope \cite{eht_papers} confirmed the success of Einstein's general relativity as a classical theory of gravity. However, standard General Relativity also fails as a fundamental theory to explain gravity interaction at a quantum level. 

Therefore, there are several motivations to explore theories beyond the standard formulation of gravity. Among this effort, we highlight models based on the so-called symmetric teleparallel gravity or $f(Q)$ gravity, introduced by Jimenez et al. \cite{Jimenez/2018}, where the nonmetricity $Q$ is responsible to the gravitational interaction. Investigations on $f(Q)$ gravity have been rapidly developed as well as observational constraints to confront it against standard GR formulation. 

An interesting set of constraints on $f(Q)$ gravity were done by Lazkoz et al. \cite{Lazkoz/2019}, were the $f(Q)$ Lagrangian is written as polynomial functions of the redshift $z$. The constraints for these models were successfully derived using data from the expansion rate, Type Ia Supernovae, Quasars, Gamma-Ray Bursts, Baryon Acoustic Oscillations data, and Cosmic Microwave Background distance. Another relevant analysis about $f(Q)$ gravity consists in understand its behavior under different energy conditions. 

As it is known, the energy conditions represent paths to implement the positiveness of the stress-energy tensor in the presence of matter. Moreover, they can be used to describe the attractive nature of gravity, besides assigning the fundamental causal and the geodesic structure of space-time \cite{Capozziello/2018}. In this paper, we studied the strong, the weak, the null, and the dominant energy conditions for $f(Q)$ gravity, working with a perfect fluid matter distribution. The actual accelerating phase our Universe passes through, constraints that the strong energy condition should be violated. This constraint, together with the actual values of Hubble and deceleration parameters allowed us to test the viability of different forms of $f(Q)$ gravity. 

The ideas presented in this paper are organized in the following way: in section \ref{II} we present the generalities on $f(Q)$ gravity, the field equations as well as the energy conservation equation for a perfect fluid. In section \ref{III}, we show the explicit forms of the energy conditions derived from the Raychaudhury equations. The two scenarios for $f(Q)$ gravity and their constraints are carefully analyzed through section \ref{IV}. We also verified the deviations between our scenarios and $\Lambda$CDM cosmological model in section \ref{V}. Our final remarks and perspectives are discussed in section \ref{VI}.

\section{Motion Equations in $f(Q)$ gravity}\label{II}

Let us consider the action for $f(Q)$ gravity given by \cite{Jimenez/2018}
\begin{equation}
\label{1}
\mathcal{S}=\int \frac{1}{2}\,f(Q)\,\sqrt{-g}\,d^4x+\int \mathcal{L}_m\,\sqrt{-g}\,d^4x\,,
\end{equation}
where $f(Q)$ is a general function of the $Q$, $\mathcal{L}_m$ is the matter Lagrangian density and $g$ is the determinant of the metric $g_{\mu\nu}$.\\
The nonmetricity tensor and its traces are such that
\begin{equation}
\label{2}
Q_{\gamma\mu\nu}=\nabla_{\gamma}g_{\mu\nu}\,,
\end{equation}
\begin{equation}
\label{3}
Q_{\gamma}={{Q_{\gamma}}^{\mu}}_{\mu}\,, \qquad \widetilde{Q}_{\gamma}={Q^{\mu}}_{\gamma\mu}\,.
\end{equation}
Moreover, the superpotential as a function of nonmetricity tensor is given by
\begin{equation}
\label{4}
4{P^{\gamma}}_{\mu\nu}=-{Q^{\gamma}}_{\mu\nu}+2Q_{({\mu^{^{\gamma}}}{\nu})}-Q^{\gamma}g_{\mu\nu}-\widetilde{Q}^{\gamma}g_{\mu\nu}-\delta^{\gamma}_{{(\gamma^{^{Q}}}\nu)}\,,
\end{equation}
where the trace of nonmetricity tensor \cite{Jimenez/2018} has the form
\begin{equation}
\label{5}
Q=-Q_{\gamma\mu\nu}P^{\gamma\mu\nu}\,.
\end{equation}
Another relevant ingredient for our approach is the energy-momentum tensor for the matter, whose definition is
\begin{equation}
\label{6}
T_{\mu\nu}=-\frac{2}{\sqrt{-g}}\frac{\delta(\sqrt{-g}\mathcal{L}_m)}{\delta g^{\mu\nu}}\,.
\end{equation}
Taking the variation of action \eqref{1} with respect to metric tensor, one can find the field equations
\begin{widetext}
\begin{equation}
\label{7}
\frac{2}{\sqrt{-g}}\nabla_{\gamma}\left( \sqrt{-g}f_Q {P^{\gamma}}_{\mu\nu}\right)+\frac{1}{2}g_{\mu\nu}f
+f_Q\left(P_{\mu\gamma i}{Q_{\nu}}^{\gamma i}-2Q_{\gamma i \mu}{P^{\gamma i}}_{\nu} \right)=-T_{\mu\nu}\,,
\end{equation}
\end{widetext}
where $f_Q=\frac{df}{dQ}$. Besides, we can also take the variation of \eqref{1} with respect to the connection, yielding to 
\begin{equation}\label{8}
\nabla_{\mu}\nabla_{\gamma}\left( \sqrt{-g}f_Q {P^{\gamma}}_{\mu\nu}\right)=0\,.
\end{equation}
Here we are going to consider the standard Friedmann-Lema\^{\i}tre-Robertson-Walker (FLRW) line element, which is explicit written as
\begin{equation}
\label{9}
ds^2=-dt^2+a^2(t)\delta_{\mu\nu}dx^{\mu}dx^{\nu}\,,
\end{equation}
where $a(t)$ is the scale factor of the Universe. The previous line element enable us to write the trace of the nonmetricity tensor as
\begin{align*}
Q=6H^2\,.
\end{align*} 
Now, let us take the energy-momentum tensor for a perfect fluid, or
\begin{equation}
\label{10}
T_{\mu\nu}=(p+\rho)u_{\mu}u_{\nu}+pg_{\mu\nu}\,,
\end{equation}
where $p$ represents the pressure and $\rho$ represents the energy density. Therefore, by substituting \eqref{9}, and \eqref{10} in \eqref{7} one can find 
\begin{equation}
\label{11}
3H^2=\frac{1}{2f_Q}\left(-\rho+\frac{f}{2}\right)\,,
\end{equation}
\begin{equation}
\label{12}
\dot{H}+3H^2+\frac{\dot{f_Q}}{f_Q}H=\frac{1}{2f_Q}\left(p+\frac{f}{2}\right)\,,
\end{equation}
as the modified Friedmann equations for $f(Q)$ gravity. Here dot $(^.)$ represents one derivative with respect to time.
The modified Friedmann equations enable us to write the density and the pressure for the Universe as
\begin{equation}
\label{14}
\rho=\frac{f}{2}-6H^2f_Q\,,
\end{equation}
\begin{equation}
\label{15}
p=\left(\dot{H}+3H^2+\frac{\dot{f_Q}}{f_Q}H\right)2f_Q-\frac{f}{2}\,.
\end{equation}

In analogy with GR, we can rewrite Eq.\eqref{11}, \eqref{12} as
\begin{equation}
3H^2=-\frac{1}{2}\tilde{\rho}\,,
\end{equation}
\begin{equation}
\dot{H}+3H^2=\frac{\tilde{p}}{2}\,.
\end{equation}
where
\begin{equation}
\label{13a}
\tilde{\rho}=\frac{1}{f_Q}\left(\rho-\frac{f}{2}\right)\,,
\end{equation}
\begin{equation}
\label{14a}
\tilde{p}=-2\,\frac{\dot{f_Q}}{f_Q}\,H+\frac{1}{f_Q}\,\left(p+\frac{f}{2}\right)\,.
\end{equation}
The previous equations are going to be components of a modified energy-momentum tensor $\tilde{T}_{\,\mu\nu}$, embedding the dependence on the trace of the nonmetricity tensor.
\section{Energy conditions}\label{III}

The energy conditions (ECs) are the essential tools to understand the geodesics of the Universe. Such conditions can be derived from the well-known Raychaudhury equations, whose forms are \cite{Poisson/2004}
\begin{equation}
\label{16}
\frac{d\theta}{d\tau}=-\frac{1}{3}\theta^2-\sigma_{\mu\nu}\sigma^{\mu\nu}+\omega_{\mu\nu}\omega^{\mu\nu}-R_{\mu\nu}u^{\mu}u^{\nu}\,,
\end{equation}
\begin{equation}
\label{17}
\frac{d\theta}{d\tau}=-\frac{1}{2}\theta^2-\sigma_{\mu\nu}\sigma^{\mu\nu}+\omega_{\mu\nu}\omega^{\mu\nu}-R_{\mu\nu}n^{\mu}n^{\nu}\,,
\end{equation}
where $\theta$ is the expansion factor, $n^{\mu}$ is the null vector, and $\sigma^{\mu\nu}$ and $\omega_{\mu\nu}$ are, respectively, the shear and the rotation associated with the vector field $u^{\mu}$. For attractive gravity, equations \eqref{16}, and \eqref{17} satisfy the following conditions
\begin{align}
\label{18}
R_{\mu\nu}u^{\mu}u^{\nu}\geq0\,,\\
 R_{\mu\nu}n^{\mu}n^{\nu}\geq0\,.
\end{align}
Therefore, if we are working with a perfect fluid matter distribution, the energy conditions recovered from standard GR are
\begin{itemize}
\item Strong energy conditions (SEC) if  $\tilde{\rho}+3\tilde{p}\geq 0\,$;

\item Weak energy conditions (WEC) if  $\tilde{\rho}\geq 0, \tilde{\rho}+\tilde{p}\geq 0\,$;

\item Null energy condition (NEC) if  $\tilde{\rho}+\tilde{p}\geq 0\,$;

\item Dominant energy conditions (DEC) if $\tilde{\rho}\geq 0, |\tilde{p}|\leq \rho\,$.
\end{itemize}

Taking Eqs.  $(\ref{13a})$ and $(\ref{14a})$  into WEC, NEC and DEC constraints, we are able to prove that

\begin{itemize}
\item Weak energy conditions (WEC) if  $\rho\geq 0, \rho+p\geq 0\,$;

\item Null energy condition (NEC) if  $\rho+p\geq 0\,$;

\item Dominant energy conditions (DEC) if $\rho\geq 0, |p|\leq \rho\,$.
\end{itemize}
corroborating with the work from Capozziello et al.\cite{Capozziello/2018}. In the case of SEC condition, we yield to the constraint
\begin{equation}\label{15a}
\rho+3\,p-6\,\dot{f_Q}\,H+f \geq 0\,.
\end{equation}

Moreover, let us consider the Hubble, deceleration, jerk, and snap parameters, whose forms are
\begin{eqnarray}
\label{19}
&&
H=\frac{\dot{a}}{a}\,,\qquad q=-\frac{1}{H^2}\frac{\ddot{a}}{a}\,,\\ \nonumber
&&
j=\frac{1}{H^3}\frac{\dot{\ddot{a}}}{a}\,, \qquad s=\frac{1}{H^4}\frac{\ddot{\ddot{a}}}{a}\,.
\end{eqnarray}
Such parameters enable us to represent the time derivatives of $H$ as
\begin{equation}
\label{20}
\dot{H}=-H^2(1+q)\,,
\end{equation}
\begin{equation}
\label{21}
\ddot{H}=H^3(j+3q+2)\,,
\end{equation}
\begin{equation}
\label{22}
\dot{\ddot{H}}=H^4(s-2j-5q-3)\,.
\end{equation}
So, by using Eqs. \eqref{20}-\eqref{22}, we can rewrite \eqref{14}, and \eqref{15} as
\begin{equation}
\label{23}
\rho=\frac{f}{2}-6H^2f_Q\,,
\end{equation}
\begin{equation}
\label{24}
p=\left(-H^2(1+q)+3H^2+\frac{\dot{f_Q}}{f_Q}H\right)2f_Q-\frac{f}{2}\,,
\end{equation}
which are the density and the pressure for the $f(Q)$ gravity. Therefore, the previous equations establish the following constraints for the energy conditions:
\begin{widetext}
\begin{equation}
\label{25}
\text{\textbf{SEC :}} \ \rho+3p-6\dot{f_Q}H+f=\frac{f}{2}-6H^2f_Q+3\left(-H^2(1+q)+3H^2+\frac{\dot{f_Q}}{f_Q}H\right)2f_Q-3\frac{f}{2}-6\dot{f_Q}H+f\geq0 \,,
\end{equation}
\begin{equation}
\label{26}
\text{\textbf{NEC :}} \ \rho+p=-6H^2f_Q+\left(-H^2(1+q)+3H^2+\frac{\dot{f_Q}}{f_Q}H\right)2f_Q\geq0\, ,
\end{equation}
\begin{equation}\label{27}
\text{\textbf{WEC :}} \ \rho=\frac{f}{2}-6H^2f_Q\geq0,
\rho+p=-6H^2f_Q+\left(-H^2(1+q)+3H^2+\frac{\dot{f_Q}}{f_Q}H\right)2f_Q\geq0\, ,
\end{equation}
\begin{equation}\label{28}
\text{\textbf{DEC :}} \ \rho=\frac{f}{2}-6H^2f_Q\geq0,
\rho\pm p=\frac{f}{2}-6H^2f_Q\pm3\left(-H^2(1+q)+3H^2+\frac{\dot{f_Q}}{f_Q}H\right)2f_Q-3\frac{f}{2}\geq0\, .
\end{equation}
\end{widetext}

\section{Constraining $f(Q)$ Theories}\label{IV}

In this section, we discuss the viability of the functional form of $f(Q)$ in FLRW spacetime. In order to do so,  we take the present values for the Hubble and the decelerating parameters as $H_0=67.9\,,$ and $q_0=-0.503$, respectively \cite{Plank/2018,Capozziello/2019}. Moreover, several observations confirm that the Universe is going through an accelerated phase \cite{Riess/1998}, carried by a negative pressure regime. Such a scenario imposes that SEC needs to be violated \cite{Plank/2018}. 

There are several approaches in the literature deriving energy conditions beyond Einstein's GR, we can see for instance, EC constraints in $f(R)$ theory \cite{Santos/2007,Bertolami/2009}, $f(G)$ theory \cite{Gracia/2011,Bamba/2017}, $f(T)$ theory \cite{Liu/2012}, $f(\mathcal{G},T)$ theory \cite{Sharif/2016}, $f(R,T,R_{\mu\nu}T^{\mu\nu})$ theory \cite{Sharif/2013}, $f(R,\mathcal{G})$ theory \cite{Atazadeh/2014}, $f(R,\square R,T)$ gravity \cite{Yousaf/2018}, $f(R,T)$ theory \cite{Moraes/2019} etc. However, the previous studies are mainly focused on the WEC energy condition, whereas our intent in this paper is to study the constraint of all the ECs in $f(Q)$ theory. To investigate the ECs with the present values of the geometrical parameters in $f(Q)$ theory, we need to fix the functional form of $f(Q)$. Once this form fixed, it will be easy to investigate the cosmological scenarios. In their beautiful work, T. Harko, et al. \cite{Harko/2018} discussed the coupling matter in modified $Q$ gravity by assuming a power-law and an exponential form of $f(Q)$.  This investigation, motivated us to work with a polynomial form for $f(Q)$ gravity. Moreover, we also introduce a logarithmic dependence of $f(Q)$, which we are going to analyze carefully.

\subsection{$f(Q)= Q+mQ^n$}

\begin{figure}[H]
\subfloat[Energy density\label{sfig:testa}]{
  \includegraphics[scale =0.26]{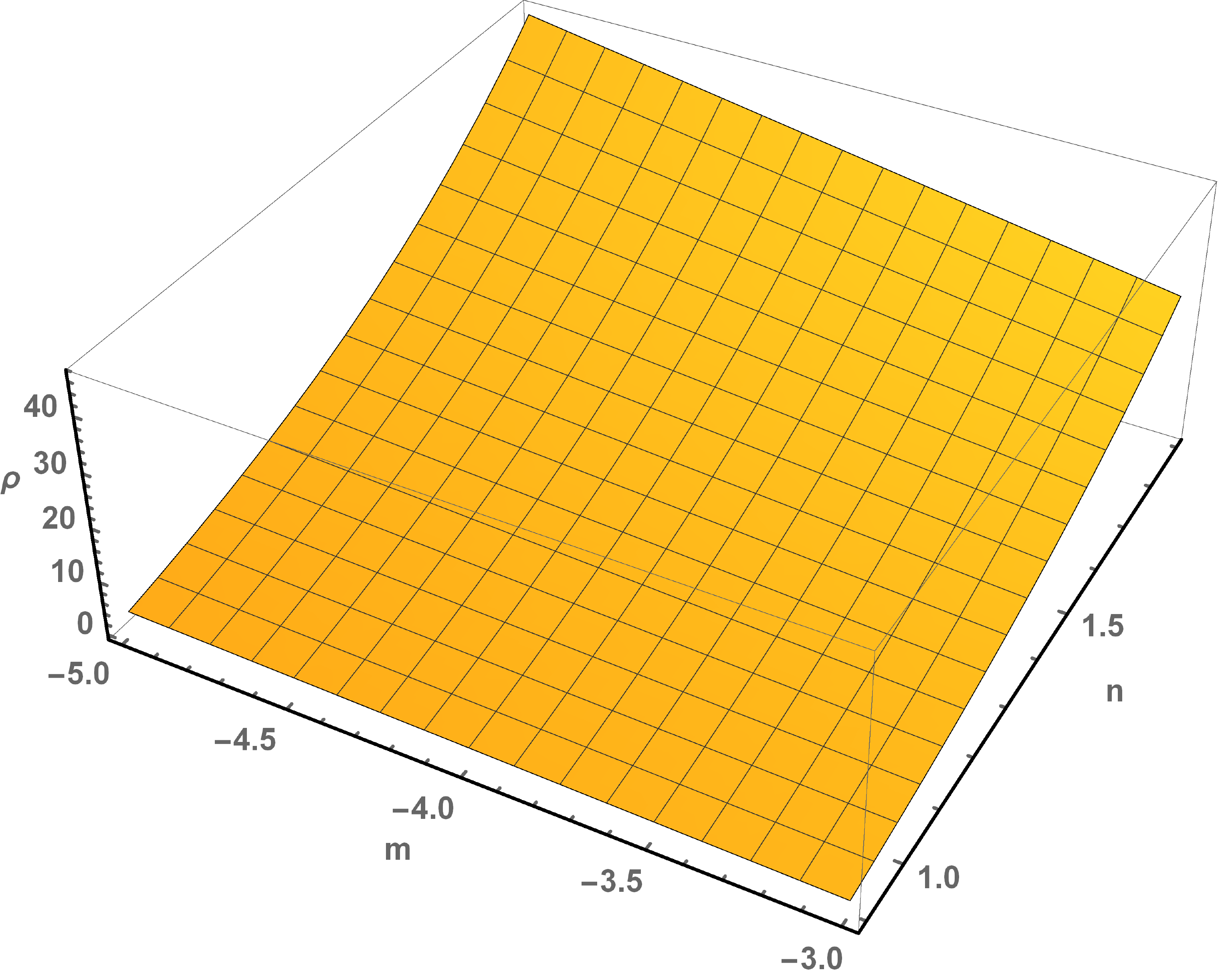}
}
\subfloat[SEC\label{sfig:testa}]{
  \includegraphics[scale =0.25]{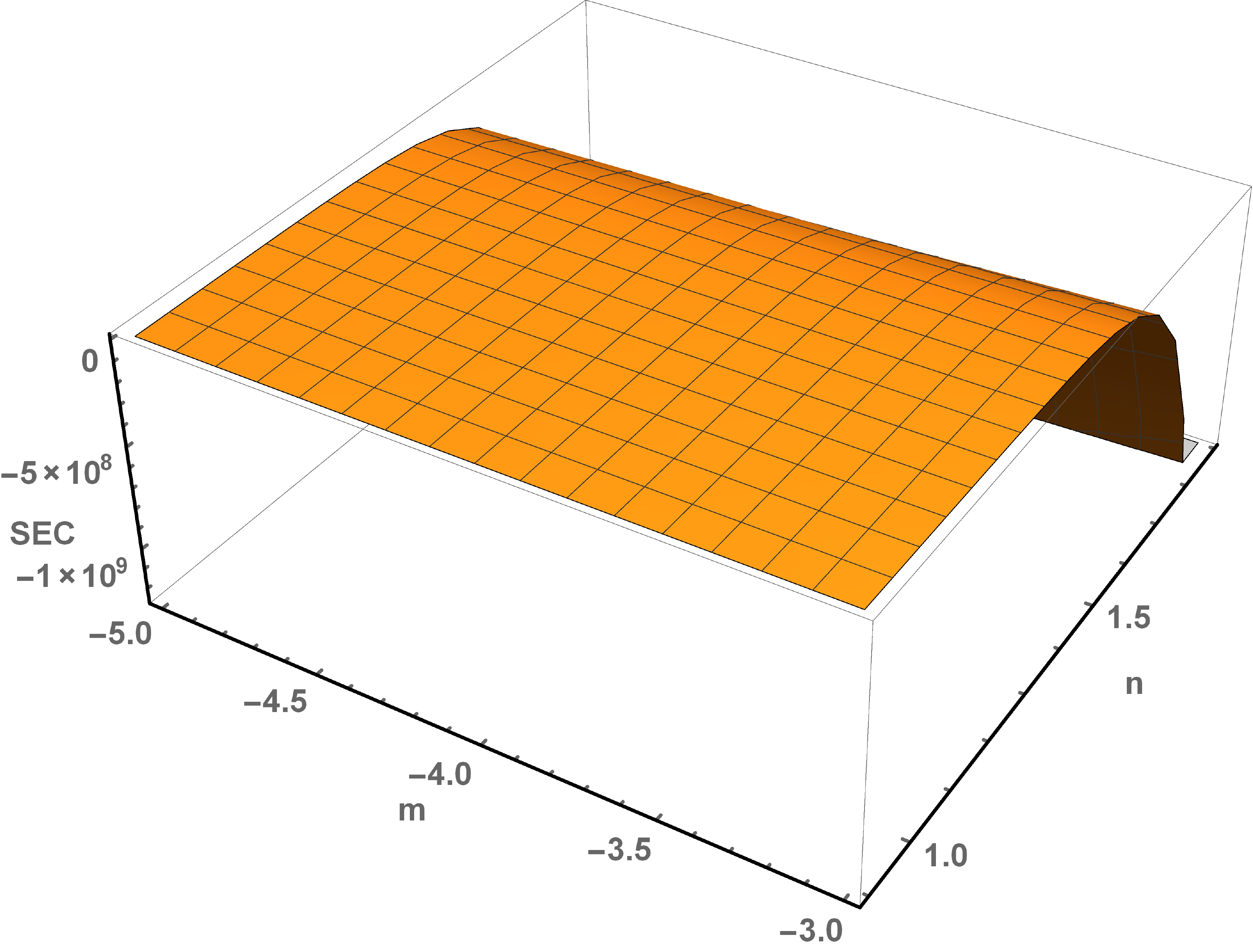}
}\hfill
\subfloat[WEC\label{sfig:testa}]{
  \includegraphics[scale =0.26]{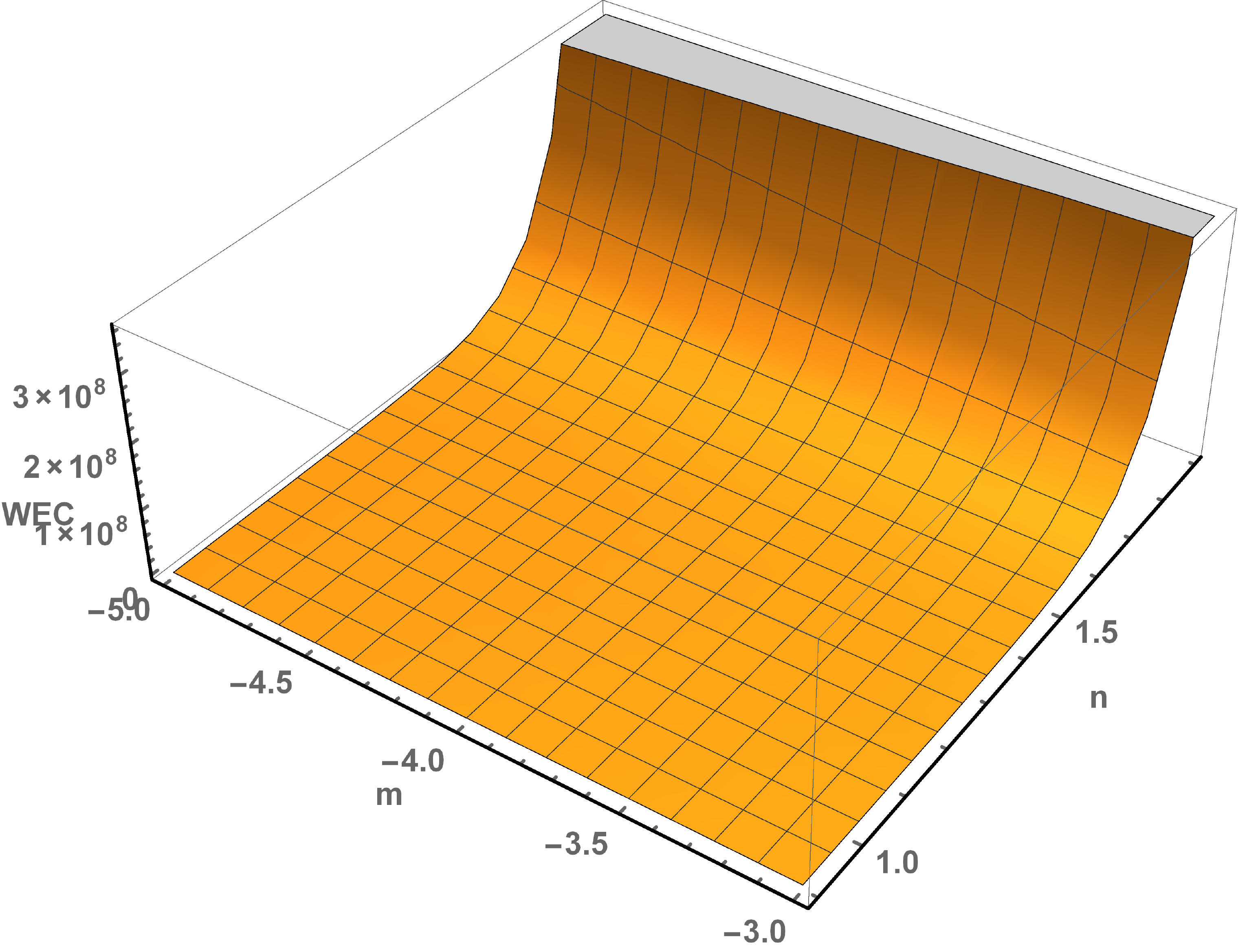}
}
\subfloat[DEC\label{sfig:testa}]{
  \includegraphics[scale =0.26]{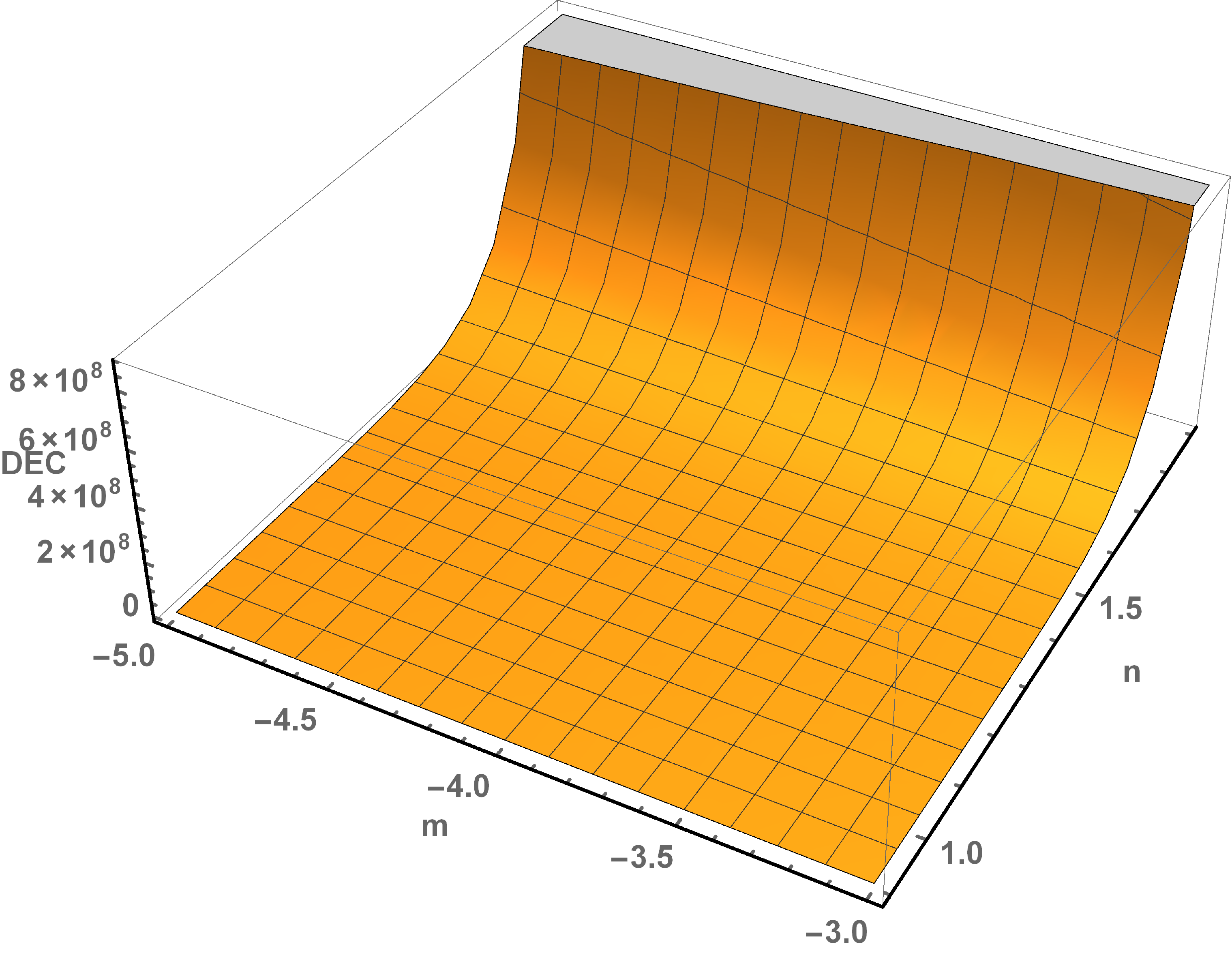}
}
\caption{Energy conditions for $f(Q)= Q+mQ^n$ derived with the present values of $H_0$ and $q_0$ parameters.}
\label{f1}
\end{figure}

In this subsection, we presume the $f(Q)$ as a algebraic polynomial function of $Q$ with free parameters $m$ and $n$. The previous function establishes that the ECs need to satisfy the following conditions:
 \begin{widetext}
\begin{multline}\label{29}
\text{\textbf{SEC : }}3 H_0^2-m 6^n \left\lbrace 2^{-1} (2 n-1)-1\right\rbrace H_0^{2n}+\frac{1}{2} \left[H_0^2 (6-12 q_0)-m 6^n (2 n-1) H_0^{2n} \left\lbrace n (q_0+1)-3\right\rbrace\right]+ \\
 2^{1 + n} 3^n H_0^{2n} m (-1 + n) n (1 + q_0)\geq0\,,
\end{multline}
\begin{equation}\label{30}
\text{\textbf{NEC : }}-3 H_0^2-m2^{-1}6^n (2 n-1) H_0^{2n}+\frac{1}{6} \left[H_0^2 (6-12 q_0)-m 6^n (2 n-1) H_0^{2n} \left\lbrace n (q_0+1)-3\right\rbrace\right]\geq0\,,
\end{equation}
\begin{multline}\label{31}
\text{\textbf{WEC : }}-3 H_0^2-m2^{-1}6^n (2 n-1) H_0^{2n}\geq0,\\ \text{and }
-3 H_0^2-m2^{-1}6^n (2 n-1) H_0^{2n}+\frac{1}{6} \left[H_0^2 (6-12 q_0)-m 6^n (2 n-1) H_0^{2n} \left\lbrace n (q_0+1)-3\right\rbrace\right]\geq0\,,
\end{multline}
\begin{multline}\label{32}
\text{\textbf{DEC : }}-3 H_0^2-m2^{-1}6^n (2 n-1) H_0^{2n}\geq0,\\ \text{and }
-3 H_0^2-m2^{-1}6^n (2 n-1) H_0^{2n}\pm\frac{1}{6} \left[H_0^2 (6-12 q_0)-m 6^n (2 n-1) H_0^{2n} \left\lbrace n (q_0+1)-3\right\rbrace\right]\geq0\,.
\end{multline}
\end{widetext}
From \eqref{29}-\eqref{32}, one can easily observe that the ECs directly depend on the free parameters $m$ and $n$. Nevertheless, one cannot take the values of $m$ and $n$ arbitrarily, which may violate the ECs as well as the current scenario of the Universe dominated by the dark energy. Therefore, using \eqref{29}-\eqref{32}, we found some restrictions on the model parameters $m$ and $n$. Using WEC, we found that $m$ should be less than or equal to $-1$ ($m\leq -1$), and $n$ should be greater than or equal to $0.9$ ($n\geq 0.9$). Also, we found that \eqref{29}, \eqref{30}, \eqref{32} reduces the range of the model parameter to $0.9\leq n\leq 2$, in order to proper describe SEC, NEC and DEC. Finally, we conclude that for $m\leq -1$ and $0.9\leq n\leq 2$, this model represents the current stage of the Universe. In addition to this, we showed the profile of all energy conditions for some range of model parameters $m$, and $n$. From Fig. \ref{f1}, one can observe that WEC, NEC, DEC are satisfied, while SEC is violated, corroborating with an accelerated expansion \cite{Visser/2000,Moraes/2017}.

%
%
%

\subsection{$f(Q)=\alpha+\beta\log Q$}

Here, we introduce $f(Q)$ as a logarithmic function of the nonmetricity with free parameters $\alpha$, and $\beta$. Therefore, the ECs are impelled to obey the constraints
\begin{widetext}
\begin{equation}\label{33}
\text{\textbf{SEC : }}-\beta-2 \beta  (q_0+1) +\frac{3}{2} \left[\alpha +\beta  \log \left(6 H_0^2\right)\right]-\frac{3 H_0^2 \left[\alpha -2 \beta +\beta  \log \left(6 H_0^2\right)\right]-2 \beta  H_0^2 (q_0+1)}{2 H_0^2}\geq 0\,,
\end{equation}
\begin{equation}\label{34}
\text{\textbf{NEC : }}-\beta +\frac{1}{2} \left[\alpha +\beta  \log \left(6 H_0^2\right)\right]-\frac{3 H_0^2 \left[\alpha -2 \beta +\beta  \log \left(6 H_0^2\right)\right]-2 \beta  H_0^2 (q_0+1)}{6 H_0^2}\geq0\,,
\end{equation}
\begin{multline}\label{35}
\text{\textbf{WEC : }}-\beta+\frac{1}{2} \left[\alpha +\beta  \log \left(6 H_0^2\right)\right]\geq0,\\ \text{and }
-\beta +\frac{1}{2} \left[\alpha +\beta  \log \left(6 H_0^2\right)\right]-\frac{3 H_0^2 \left[\alpha -2 \beta +\beta  \log \left(6 H_0^2\right)\right]-2 \beta  H_0^2 (q_0+1)}{6 H_0^2}\geq0\,,
\end{multline}
\begin{multline}\label{36}
\text{\textbf{DEC : }}-\beta+\frac{1}{2} \left[\alpha +\beta  \log \left(6 H_0^2\right)\right]\geq0,\\  \text{and }
-\beta +\frac{1}{2} \left[\alpha +\beta  \log \left(6 H_0^2\right)\right]\pm\frac{3 H_0^2 \left[\alpha -2 \beta +\beta  \log \left(6 H_0^2\right)\right]-2 \beta  H_0^2 (q_0+1)}{6 H_0^2}\geq0\,.
\end{multline}
\end{widetext}

\begin{figure}[ ]
\subfloat[Energy density\label{sfig:testa}]{
  \includegraphics[scale =0.24]{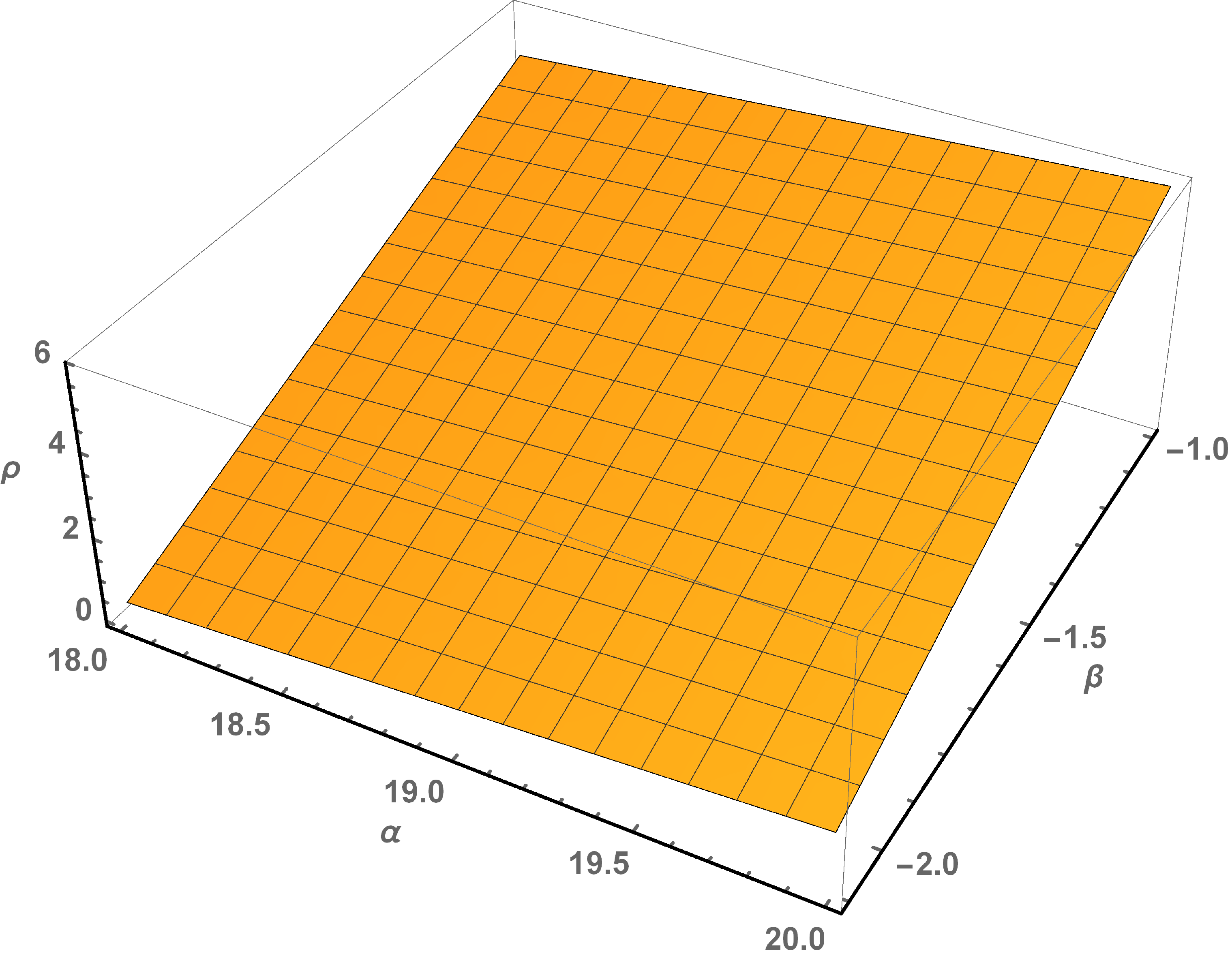}
}\hfill
\subfloat[SEC\label{sfig:testa}]{
  \includegraphics[scale =0.24]{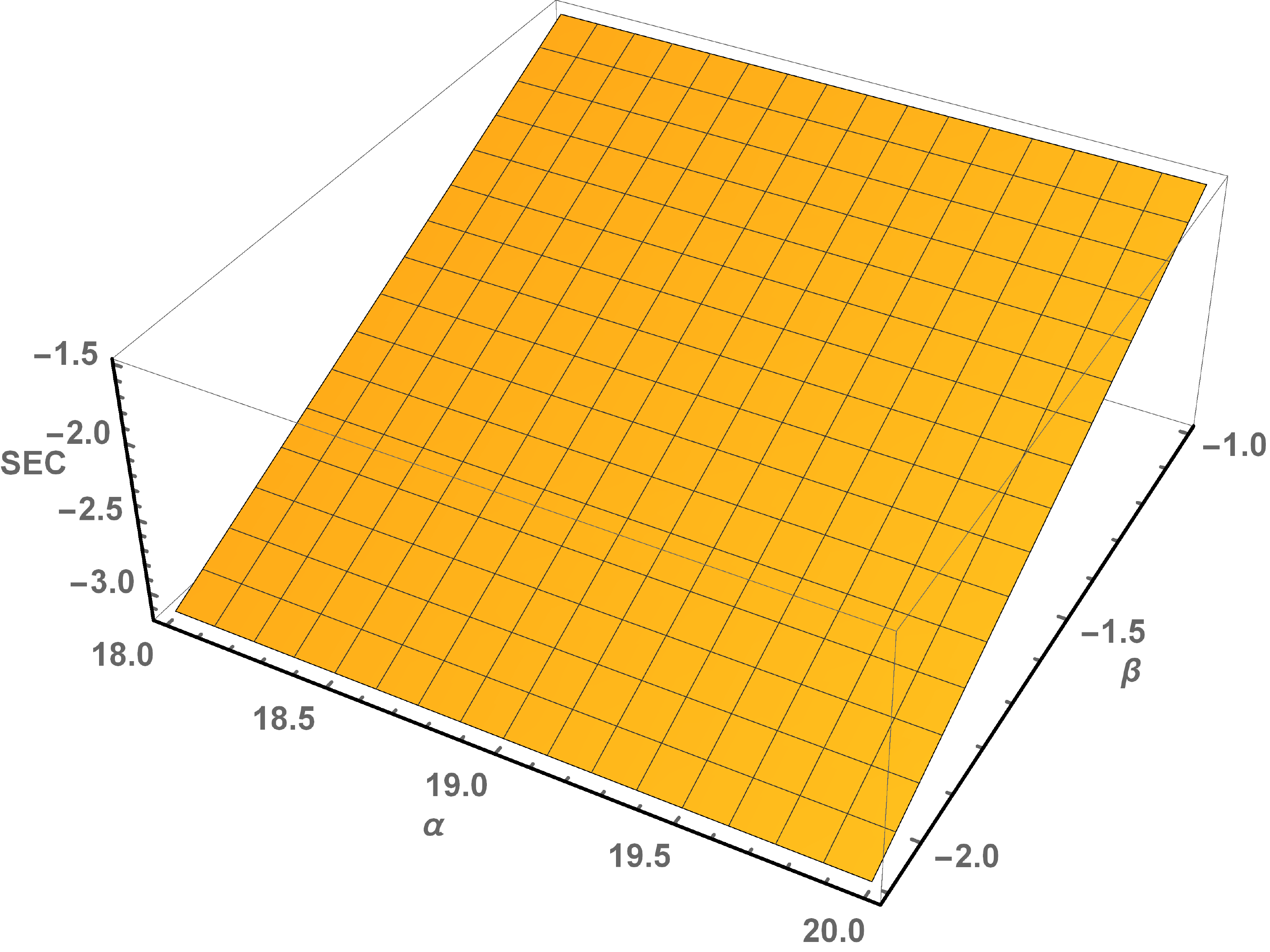}
}\hfill
\subfloat[WEC\label{sfig:testa}]{
  \includegraphics[scale =0.24]{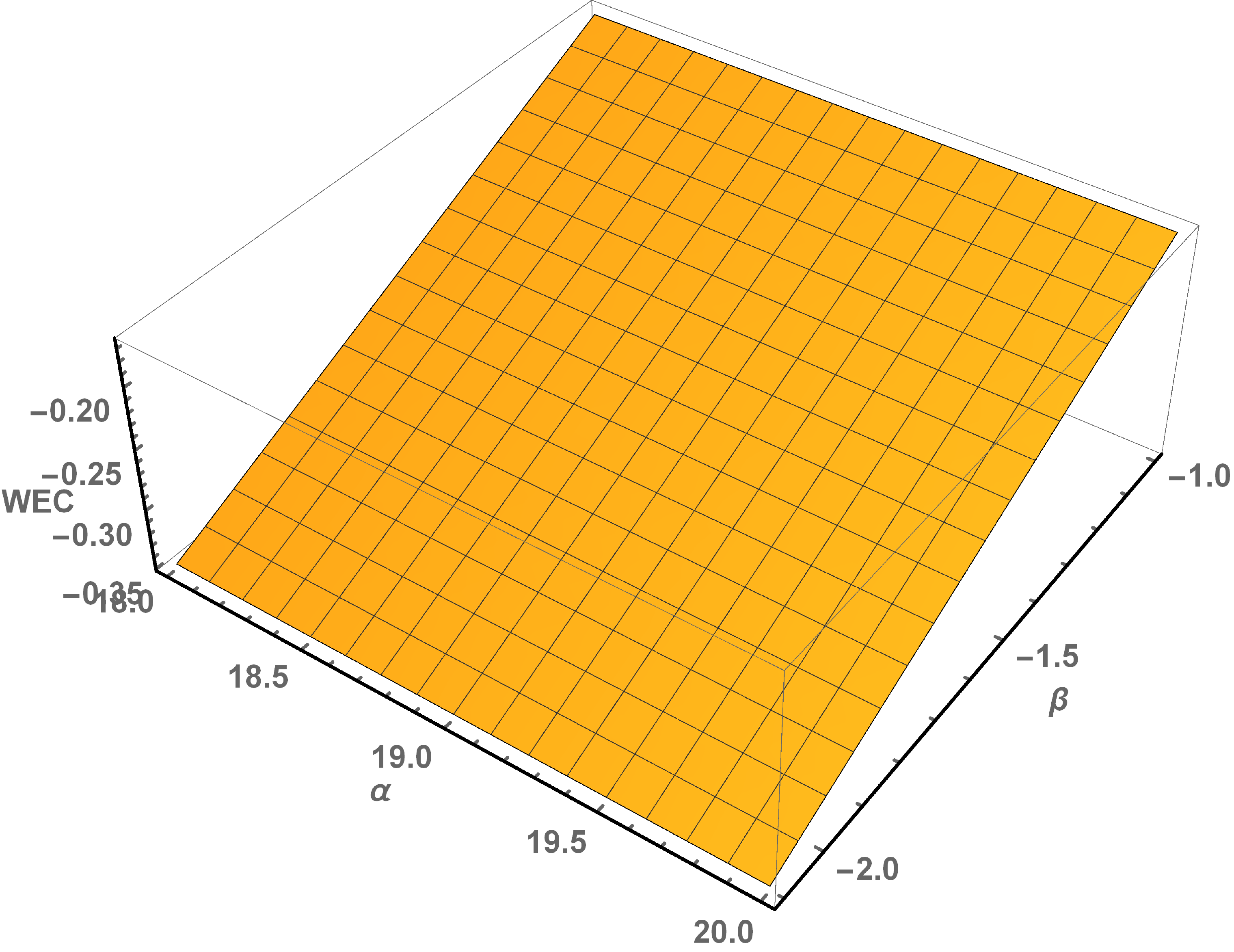}
}\hfill
\subfloat[DEC\label{sfig:testa}]{
  \includegraphics[scale =0.24]{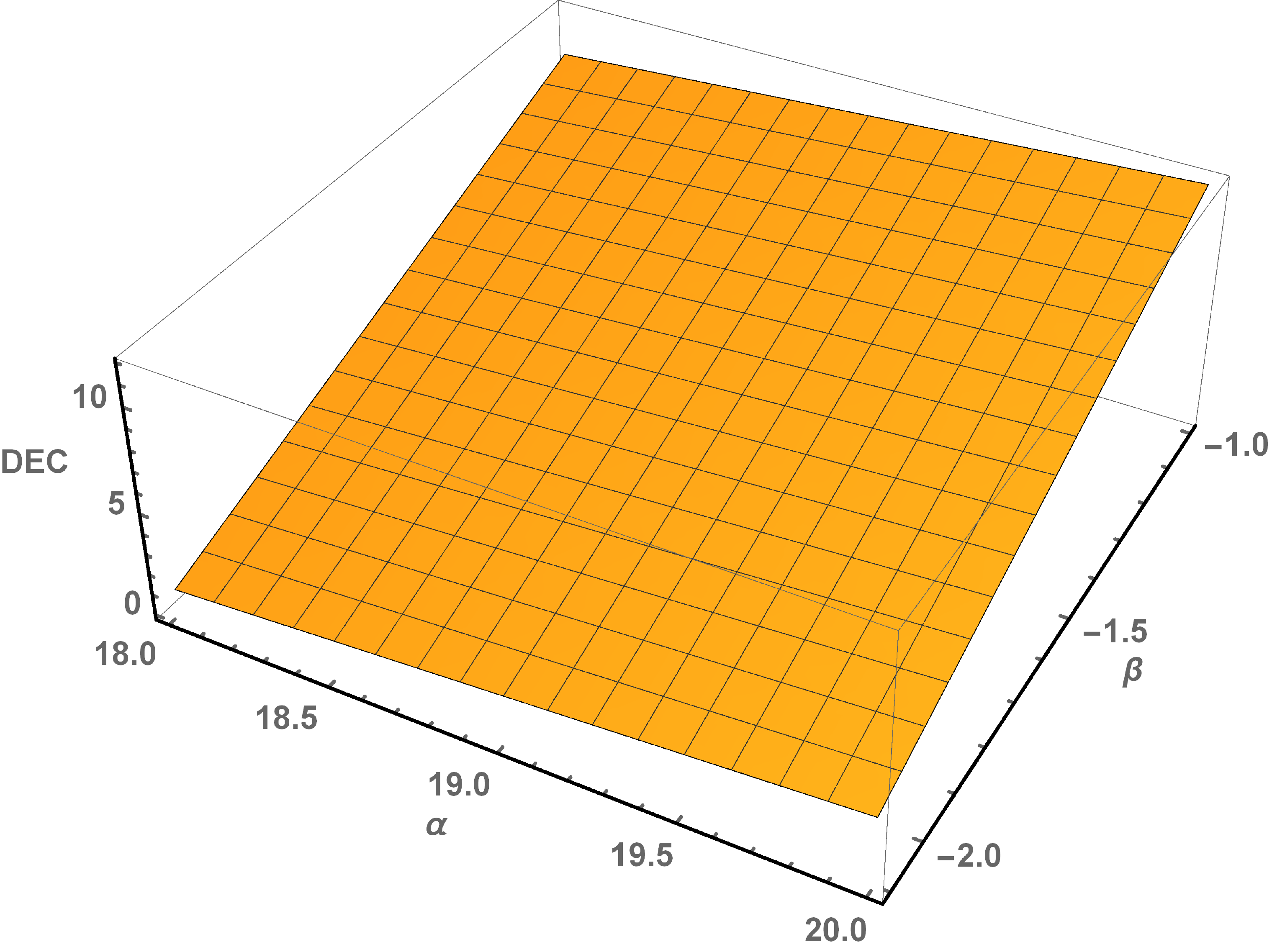}
}
\caption{Energy conditions for $f(Q)=\alpha+\beta\log Q$ derived with the present values for $H_0$, and $q_0$  parameters.}
\label{f2}
\end{figure}

The ECs showed in Eqs. \eqref{33}-\eqref{36} unveil their direct dependence on free parameters $\alpha$, and $\beta$.  The previous equations also established that we cannot choose arbitrary values for these free parameters, as observed in our previous model. Through Eqs. \eqref{33}, \eqref{34}, \eqref{35}, and \eqref{36}, we found that condition SEC is violated and condition WEC is partially satisfied ($\rho>0$) if $\alpha\geq -9\beta, (\beta\leq -1)$, besides NEC, and DEC are still obeyed. This violation of WEC with positive density notably makes this $f(Q)$  theory behaves like scalar-tensor gravity models \cite{Whinnett/2004}, and such a violation can be interpreted as natural contributions from quantum effects to classical gravity \cite{Calcagni_book}. The features of these conditions can be appreciated in Fig. \ref{f2}, where the graphics were depicted considering a specific range of values for parameters $\alpha$, and $\beta$.

\section{Deviation from the standard $\Lambda$CDM model}\label{V}

So far, the $\Lambda$CDM is the most well-succeeded model used to describe the actual observations of the Universe. Such a model has been broadly tested by several different surveys along the past few years, such as WMAP, Planck, and The Dark Energy Survey (DES). As pointed by Lazkoz et al. \cite{Lazkoz/2019}, the $f(Q)$ model can mimics the $\Lambda$CDM one by taking $f(Q)=f_{\Lambda}(Q)=-Q$. Therefore, considering this specific mapping for $f_{\Lambda}$, we find the following energy conditions 
\begin{itemize}
\item \textbf{SEC : }$6H^2 (q-1) \geq 0$,
\item \textbf{NEC : }$2H^2(1+q)\geq 0$,
\item \textbf{WEC : }$3H^2\geq 0$ and  $2H^2(1+q)\geq 0$,
\item \textbf{DEC : }$3H^2\geq 0$ and  $2H^2(1+q)\geq 0$ or, $-2H^2(-2+q)\geq 0$.
\end{itemize}

By taking the actual values of $H_0$ and $q_0$ in the above conditions, one can prove that WEC, NEC, DEC are satisfied, however, SEC condition is violated. This is the expected behavior for a standard accelerated phase for the Universe. Moreover, we realize an analogous description in respect to energy conditions between our first model for $f(Q)$ gravity, and the $\Lambda$CDM.

\begin{figure}[H]
\includegraphics[width=7.5 cm]{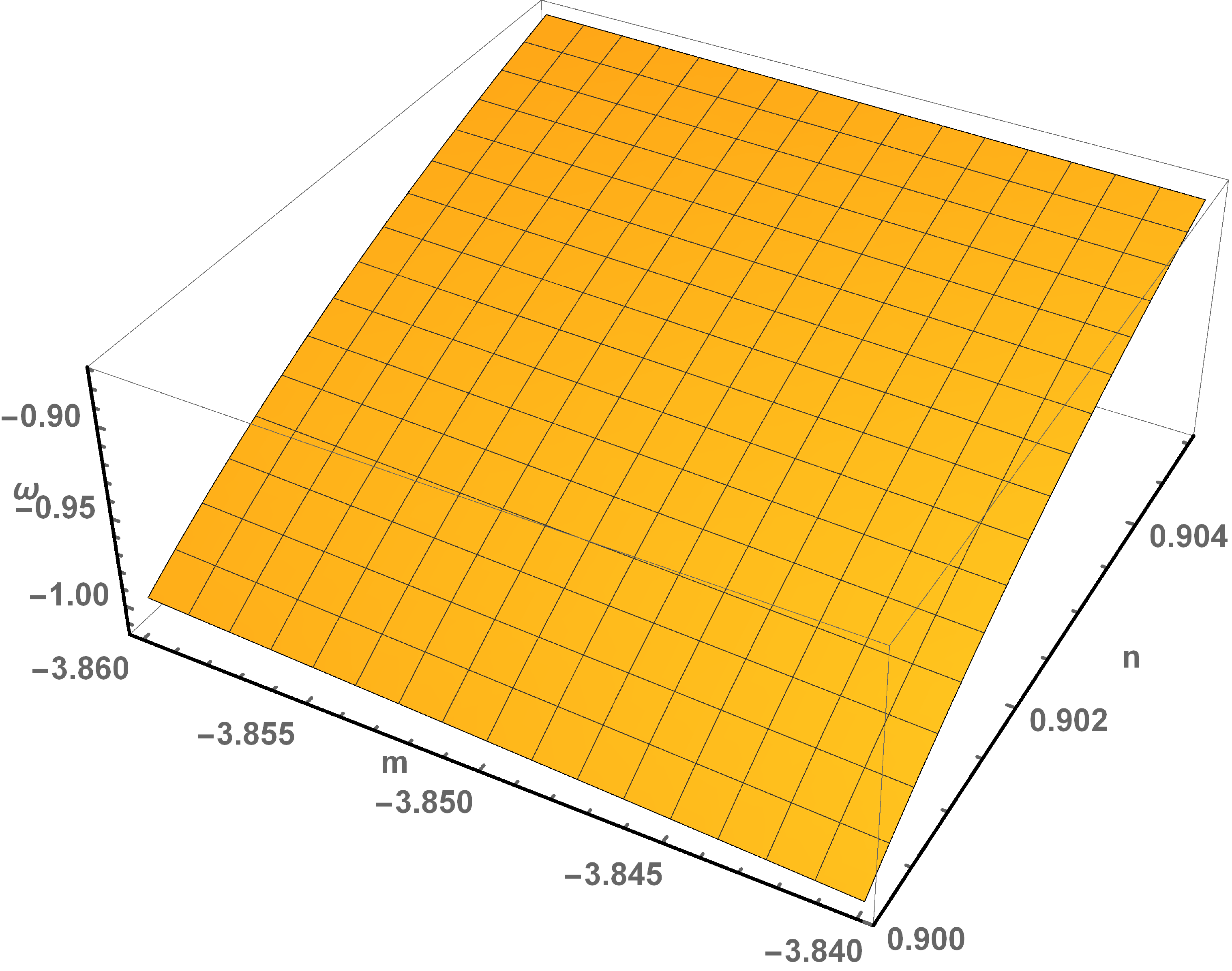}
\caption{Equation of state parameter $\omega$ for $f(Q)=Q+mQ^n$ derived with the present values for $H_0$, and $q_0$  parameters.}
\label{f3}
\end{figure}

\begin{figure}[H]
\includegraphics[width=7.5 cm]{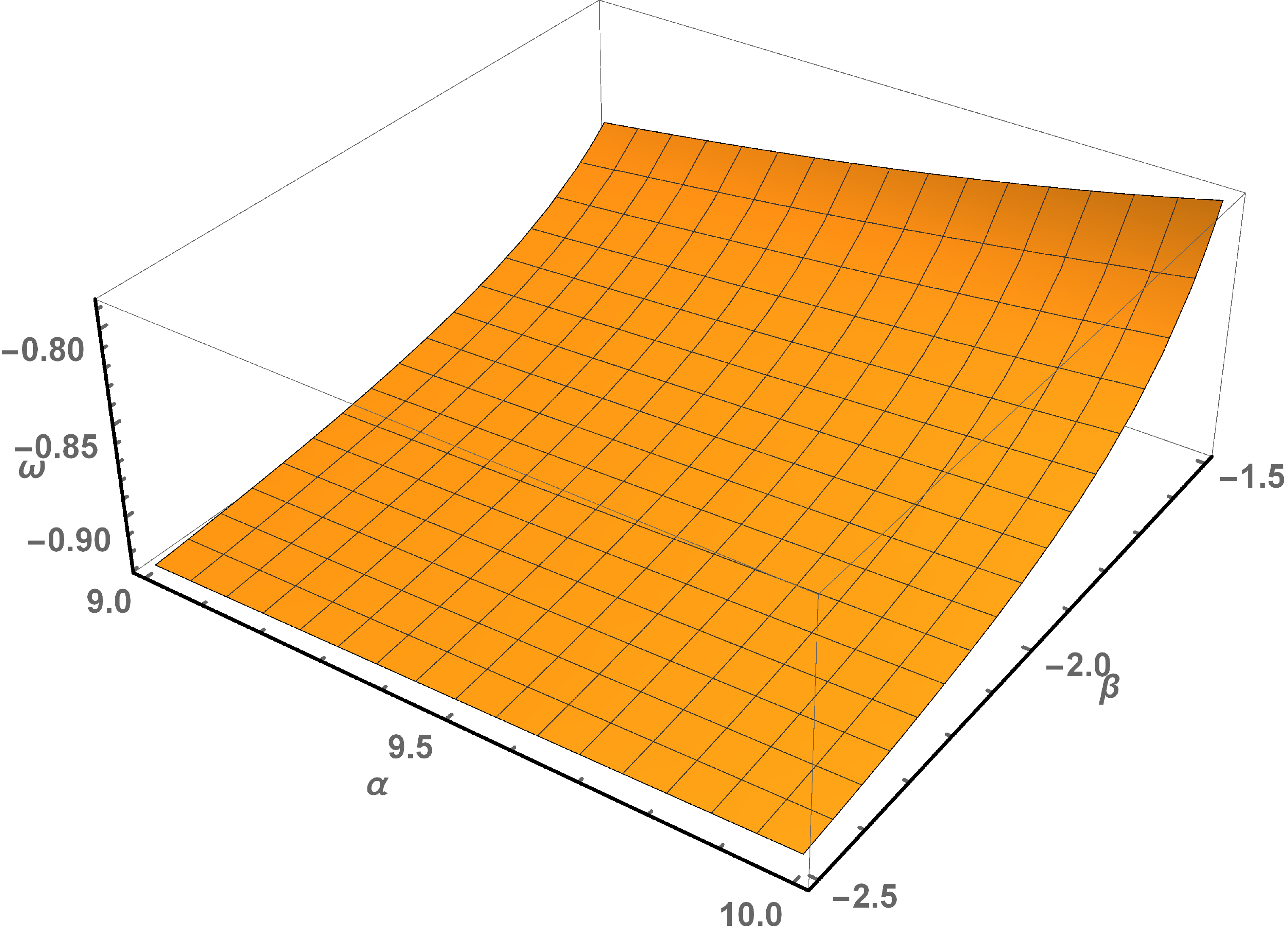}
\caption{Equation of state parameter $\omega$ for $f(Q)=\alpha+\beta\log Q$ derived with the present values for $H_0$, and $q_0$  parameters.}
\label{f4}
\end{figure}

Besides, the recent observations from Planck Collaboration, as well as the $\Lambda$CDM model, confirm that the equation of state parameter is $\omega\simeq -1$ \cite{Plank/2018}. This behavior corresponds to a negative pressure regime for Universe, which configures its current accelerated phase. Therefore, $\omega$ parameter presents as a suitable candidate to compare our models with $\Lambda$CDM. Once the equation of state parameter is defined as
\begin{equation}
\omega =\frac{p}{\rho}\,,
\end{equation}
our previous models yields to the following forms of $\omega$:
\begin{equation}
\label{37}
 \omega=-1+\frac{2 (q_0+1) \left\lbrace m 6^n n (2 n-1) H_0^{2n}+6 H_0^2\right\rbrace}{3 m 6^n (2 n-1) H_0^{2n}+18 H_0^2}\,,
\end{equation}
for $f(Q)=Q+mQ^n$ gravity, and
\begin{equation}
\label{38}
 \omega=-1+\frac{2 \beta  (q_0+1)}{3\alpha -6 \beta +3\beta  \log \left(6 H_0^2\right)}
\end{equation}
for $f(Q)=\alpha+\beta\log Q$. In Fig. \ref{f3} \& Fig. \ref{f4}, we have shown the profiles of the equation of state parameter for both $f(Q)$ models here introduced. The graphics were depicted considering the energy conditions constraints for free parameters $m$, $n$, $\alpha$, and $\beta$. From these figures, one can observe that the values of $\omega$ are very close to $-1$, which is in agreement with the recent observational data. Consequently, our models fit the equation of state parameter as good as $\Lambda$CDM, corroborating with the violation of SEC, and confirming their viability to describe an accelerated Universe. 


\section{Conclusion}\label{VI}

There are several theories of gravity beyond Einstein's GR, however, one critical role to define their self-consistencies is the energy condition. The physical motivation for a new theory of gravity is related to its compatibility with the causal and geodesic structure of space-time, which can be addressed through different sets of energy conditions. In the present study, we derived the strong, the weak, the null, and the dominant energy conditions for two different $f(Q)$ gravity models. Inspired by the work of Harko et al. \cite{Harko/2018}, our first model was a polynomial function of the nonmetricity $Q$ and has two free parameters $m$, and $n$. The energy conditions established $m\leq -1$, and $0.9\leq n\leq 2$ as constraints to describe an accelerated expansion of the Universe. 

As a second approach, we introduced a gravity model with a logarithmic dependence on the nonmetricity. Such a model means that the $f(Q)$ smoothly tends to the Einstein-Hilbert model ($f(Q) \varpropto Q$), and had two free parameters named $\alpha$ and $\beta$. The graphics presented in Fig. \ref{f2} unveil a desired accelerating Universe for $18\leq \alpha \leq 20$, and $-2\leq \beta \leq -1$. Moreover, such a theory violates both SEC and WEC with positive density, exhibiting a behavior analogous to scalar-tensor field gravity models \cite{Whinnett/2004}. 

As a matter of completeness, we compared our energy constraints with those from the $\Lambda$CDM model. In the $\Lambda$CDM gravity, all energy conditions are satisfied except SEC. This behavior is compatible with our first proposed model where $f(Q)=Q+mQ^n$, strengthening its potential as a promising new description for gravity.    

Moreover, the equation of state parameters, derived from our two $f(Q)$ approaches, are compatible with a current phase of negative pressure, presenting values close to $-1$. This behavior also corroborates with $\Lambda$CDM description for dark energy, as well as with current experimental observations \cite{Plank/2018}. 

These previous results allowed us to verify the viability of different families of $f(Q)$ gravity models, lighting new routes for a complete description of gravity compatible with the dark energy era. Another interesting fact is that there is plenty of freedom for our free parameters, enabling several testable scenarios for $f(Q)$ gravity. Such tests could include the absence of ghosts modes, gravitational waves constraints, and cosmological parameters derived from Cosmic Microwave Background. Besides, it would be interesting to study carefully the coupling of $f(Q)$ with inflaton fields, looking for possible analytic models or for cosmological parameters constraints. We intend to address some of these investigations in the near future and hope to report on them. 

\acknowledgments S.M. acknowledges Department of Science \& Technology (DST), Govt. of India, New Delhi, for awarding Junior Research Fellowship (File No. DST/INSPIRE Fellowship/2018/IF180676). JRLS would like to thank CNPq (Grant no. 420479/2018-0), CAPES, and PRONEX/CNPq/FAPESQ-PB (Grant nos. 165/2018, and 0015/2019) for financial support. We are very much grateful to the honorable referee and the editor for the illuminating suggestions that have significantly improved our work in terms of research quality and presentation.


\begin{thebibliography}{90}
\bibitem{Capozziello_book} S. Capozziello, and V. Faraoni, \textit{Beyond Einstein Gravity: A Survey of Gravitational Theories for Cosmology and Astrophysics}, Springer Dordrecht, 2011. doi: 10.1007/978-94-007-0165-6.
\bibitem{lv_papers} B. P. Abbott et al. (LIGO Scientific Collaboration and Virgo Collaboration), {\it Phys. Rev. Lett.}, {\bf 116} (2016) 061102; B. P. Abbott et al. (LIGO Scientific Collaboration and Virgo Collaboration),{\it Phys. Rev. Lett.},  {\bf 119} (2017) 161101; B. P. Abbott et al. (LIGO Scientific Collaboration and Virgo Collaboration), {\it Phys. Rev. Lett.},  {\bf 123} (2019) 011102.
\bibitem{eht_papers} The Event Horizon Telescope Collaboration et al., {\it ApJL},  {\bf 875} (2019) L1; The Event Horizon Telescope Collaboration et al., {\it ApJL},  {\bf 875} (2019) L5. 
\bibitem{Jimenez/2018} J. B. Jim\'enez, L. Heisenberg, T. Koivisto, \prd, 
\textbf{98} (2018) 044048.
\bibitem{Lazkoz/2019} R. Lazkoz et al.,  \prd, {\bf 100} (2019) 104027. 
\bibitem{Capozziello/2018} S. Capozziello et al., {\it Phys. Lett. B}, {\bf 781}  (2018) 99.
\bibitem{Poisson/2004} A. Raychaudhuri, \textit{Phys. Rev.}, \textbf{98} (1955) 1123; J.~Ehlers, \ijmpd,  \textbf{15} (2006) 1573; S. Nojiri, S. D. Odintsov, \textit{Int. J. Geom. Methods Mod. Phys.}, \textbf{4} (2007) 115.
\bibitem{Plank/2018} Planck Collaboration, arXiv:1807.06209.
\bibitem{Capozziello/2019} S. Capozziello et al., \ijmpd, \textbf{28} (2019) 1930016. 
\bibitem{Riess/1998} A. G. Riess, et al., \aj, \textbf{116} (1998) 1009; S. Perlmutter, et al., \aj, \textbf{517} (1999) 565; S. Perlmutter, et al., \aj, \textbf{598} (2003) 102S; S. Cole, et al., \mnras, \textbf{362} (2005) 505; P. A. R. Ade, et al., \aap, \textbf{571} (2014) A16. 
\bibitem{Santos/2007} J. Santos, et al., \prd, \textbf{76}  (2007) 083513.
\bibitem{Bertolami/2009} O. Bertolami, M. C. Sequeira, \prd, \textbf{79}  (2009) 104010.
\bibitem{Gracia/2011} N. M. Gracia et al., \prd, \textbf{83}  (2011) 104032.
\bibitem{Bamba/2017} K Bamba et al., \grg,  \textbf{49} (2017) 112.
\bibitem{Liu/2012} D. Liu, M. J. Rebouc,  \prd, \textbf{86} (2012) 083515.
\bibitem{Sharif/2016} M. Sharif, A. Ikram, \epjc, \textbf{76} (2016) 640.
\bibitem{Sharif/2013} M. Sharif, M. Zubair, \textit{J. High Energy Physics}, \textbf{2013}, (2013) 79.
\bibitem{Atazadeh/2014} K. Atazadeh, F. Darabi, \grg,  \textbf{46} (2014) 1664.
\bibitem{Yousaf/2018} Z. Yousaf et al. \textit{Int. J. Geom. Methods Mod. Phys.},  \textbf{15} (2018) 1850146.
\bibitem{Moraes/2019} P.H.R.S. Moraes et al. \textit{Adv. Astron.},  \textbf{2019} (2019) 8574798.
\bibitem{Harko/2018}T. Harko, et al., \prd, \textbf{98} (2018) 084043. 
\bibitem{Visser/2000} M. Visser, C. Barcelo, \textit{COSMO-99},  (2000) 98.
\bibitem{Moraes/2017} P.H.R.S. Moraes, P.K. Sahoo, \epjc, \textbf{77} (2017) 480.
\bibitem{Whinnett/2004} A. W. Whinnett, D. F. Torres, \textit{The Astrophysical Journal}, {\bf 603} (2004) L133.
\bibitem{Calcagni_book} G. Calcagni, \textit{Classical and Quantum Cosmology}, Springer Switzerland, 2017. doi: 10.1007/978-3-319-41127-9.
\end{thebibliography}
\end{document}